\documentclass[doublecol,figures]{epl2}
\title{Magnonic momentum transfer force on domain walls confined in space}
\author{D. Wang\inst{1}\thanks{E-mail: \email{dwwang@nudt.edu.cn}} \and Xi-guang Wang\inst{2} \and
Guang-hua Guo\inst{2}\thanks{E-mail: \email{guogh@mail.csu.edu.cn}}}
\shortauthor{D. Wang \etal}
\institute{
\inst{1} Department of Physics, National University of Defense Technology, Changsha 410073, Hunan, P. R. China\\
\inst{2} School of Physics Science and Technology, Central South University, Changsha 410083, Hunan, P. R. China
}
\pacs{75.30.Ds}{Spin waves}
\pacs{75.60.Ch}{Domain walls and domain structure}
\abstract{Momentum transfer from incoming magnons to a Bloch domain wall is calculated using one dimensional continuum micromagnetic analysis. Due to the confinement of the wall in space, the dispersion relation of magnons is different from that of a single domain. This mismatch of dispersion relations can result in reflection of magnons upon incidence on the domain wall, whose direct consequence is a transfer of momentum between magnons and the domain wall. The corresponding counteraction force exerted on the wall can be used for the control of domain wall motion through magnonic linear momentum transfer, in analogy with the spin transfer torque induced by magnonic angular momentum transfer.}
\begin{document}
\maketitle

Ever since its first proposal, manipulation of domain wall motion (DWM) using means other than the conventional magnetic field becomes one focus of the current research in nanomagnetism. The stimulus behind such intense investigation to find new ways for control of DWM is obvious, given the potential application of magnetic domain walls (DWs) as logic elements \cite{Allwood05} and information storage bits \cite{Parkin08}. Hydromagnetic drag \cite{Berger78}, spin transfer torque (STT) \cite{Slonczewski96,Berger96} and momentum transfer (MT) force \cite{Tatara} due to the flow of electrons in metallic materials were proposed, and STT driven DWM was already demonstrated experimentally \cite{Yamaguchi04,Klaeui05}. In contrast, using another important elementary excitation, magnons, to affect the motion of DW comes into horizon only recently. Magnon mediated electric current drag in a normal metal/ferromagnetic insulator/normal metal structure was studied by considering the interplay between electrical current and magnon current \cite{Zhang12}. Magnonic STT was demonstrated, employing micromagnetic simulation and analytical calculation \cite{Yan11}. The main difference, or advantage, of magnonic STT compared to the conventional electronic STT lies on the fact that it is mediated by the flow of magnons, rather than electrons. Hence, in contrast to conventional STT whose operation requires the use of ferromagnetic metals, magnonic STT can work in ferromagnetic insulators, ameliorating significantly the Joule heating problem in metals. However, the MT due to the reflection of magnons by a DW is rarely discussed, simply because magnons travel in an infinitely-extended one-dimensional (1D) wall without any reflection \cite{Winter61,Thiele73,Yan11}. The only effect of the presence of a DW is a wave-number dependent phase shift. For a 1D DW confined in space, the situation is different. Due to the fact that the magnon excitation spectrum of the DW is different from that of a single domain, magnons incident on the DW get reflected, with a frequency-sensitive coefficient of reflection. As magnons are reflected back, a counteraction force is exerted on the DW and can be used to initiate DWM \cite{XWang12}.

In the presence of pure magnonic STT, DWs will move opposite to the propagation direction of the incident spin wave (SW) \cite{Yan11}. However, the contrary was found in literature \cite{XWang12,Han09,Jamali10,Kim12}: magnons push DWs to move along with them. The counteraction force due to linear MT between magnons and DWs could be a possible explanation for this phenomenon. Kim \textit{et al}. \cite{Kim12} employed a similar argument to shed light on the observed drag of DW by magnons in micromagnetic simulation. Wang \textit{et al}. \cite{XWang12} included the counteraction force phenomenologically in a 1D model to understand micromagnetic simulation results, and found qualitative agreement between model and simulation. The purpose of the current Letter is to show analytically in one dimension that the MT from the reflected magnons to a DW can actually exert a force on the DW, hence affecting the DW's motion. The basic idea is similar to the recent discussion of magnonic heat conduction in DWs: the finite reflection probability of magnons plays an important role \cite{Yan12}. The change in the internal structure of a DW, which can also give rise to DWM \cite{Han09,Kim12}, is not considered in our 1D analysis.

\begin{figure}
\onefigure[width=0.9\linewidth]{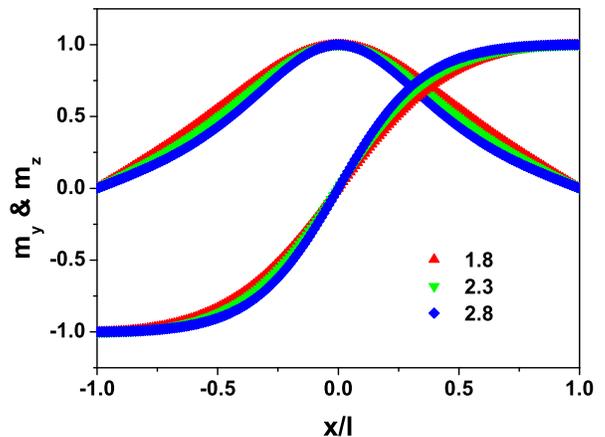}
\caption{$y$ and $z$ components of the normalized magnetization vector $\vec{m}$, $m_y$ and $m_z$, for a DW with half length $l$, whose value is given in the legend.}
\label{f1}
\end{figure}

The starting point of our discussion is the Landau-Lifshitz-Gilbert equation \cite{llg}, in the dimensionless form and omitting the damping term,
$\partial\vec{m}/\partial t= -\vec{m}\times\vec{h}$, where $\vec{m}=\vec{M}/M$ is the normalized magnetization vector, and $\vec{h}$ is the total effective field in units of the anisotropy field $\mu_0H_K=2K/M$ ($\mu_0$ is the permeability of vacuum). For a planar Bloch wall lying in the $yz$ plane (azimuthal angle $\phi=\pi/2$), the magnetization depends only on $x$ and the problem is 1D. Hence $\vec{h}=\nabla^2_x \vec{m} + m_z \hat{e}_z$, assuming that the easy anisotropy direction is along the $z$ axis and there is no external field. $\hat{e}_z$ is a unit vector pointing to the positive $z$ direction. Position $x$ and time $t$ are measured in units of the DW width $\delta=\sqrt{A/K}$ and the inverse of the spin-wave cutoff frequency $\omega_0= \gamma\mu_0 H_K$, respectively. $\gamma$ is the gyromagnetic ratio, $K$ is the uniaxial anisotropy constant and $A$ is the exchange stiffness constant. The static DW profile is determined by the LLG equation through $\vec{m}\times\vec{h} =0$. In terms of the polar angle $\theta$, $\vec{m}=(0,\sin\theta,\cos\theta)$ and the DW profile is given by (for simplicity, $\partial_x$ is used to stand for the derivative with respect to $x$, $\partial/\partial x$) $\partial_x^2\theta= \cos\theta\sin\theta$. If the DW is confined to an interval of finite length 2$l$, fixed boundary conditions are imposed as $\theta|_{x=-l}=\pi$ and $\theta|_{x=l}=0$. The corresponding solution is given by the Jacobi elliptic sine function \cite{A&S} sn with modulus $m$, $\cos\theta = \mbox{sn}(s,m)$ in terms of the scaled coordinate $s=x/\sqrt{m}$. The determination of the modulus $m$ is through the boundary conditions, which give $l=\sqrt{m}K(m)$, where $K(m)$ is the Jacobi elliptic integral of the first kind \cite{A&S}. The DW profile for various $l$ is shown in Fig. \ref{f1}. Depending on the value of $l$, $m$ varies from 0 to 1. $m=1$ corresponds to the case of an unbounded DW, $l\rightarrow\infty$, and the solution reduces to the well-known Walker profile \cite{Schryer74} $\cos\theta= \mbox{tanh}\; x$. Notice that the continuum approximation used here implies that the DW length $2l$ must be larger than the lattice constant $a$, $2l\gg a/\delta$ in dimensionless form.

For the excitation spectrum of the DW, we consider perturbations from the static equilibrium $\vec{m}=\vec{e}_r+m_\theta \vec{e}_\theta +m_\phi \vec{e}_\phi$, where $m_\theta$ and $m_\phi$ are infinitesimal deviations. For convenience, polar coordinate is used. The relationship between polar and cartesian coordinates is defined through $\vec{e}_r = \cos\theta \hat{e}_z +\sin\theta \hat{e}_y$, $\vec{e}_\phi = -\hat{e}_x$ and $\vec{e}_\theta = \vec{e}_\phi\times\vec{e}_r$. Substitute $\vec{m}$ into the LLG equation and linearize it, we get the eigenequations for the excitation amplitude
\begin{eqnarray}
\label{eigen}
i\omega m_\theta&=& (\cos2\theta-\frac{m_1}{m}-\partial_x^2)m_\phi,\nonumber\\
i\omega m_\phi&=& (\partial_x^2-\cos2\theta)m_\theta,
\end{eqnarray}
assuming a simple harmonic dependence on time for both $m_\theta$ and $m_\phi$ ($m_\theta$, $m_\phi \propto\exp(-i\omega t)$). $m_1$ is the complementary modulus, $m+m_1=1$. Eq. (\ref{eigen}) has the form of the Schr\"{o}dinger equation for a particle confined in a sn$^2$ potential formed by the DW. The corresponding propagating solution is given by \cite{W&W,Sutherland73,Musevic94}
\begin{equation}
\label{psi}
\psi=\frac{H(s+s_0)}{\Theta(s)}e^{-sZ(s_0)},
\end{equation}
where $H$, $\Theta$ and $Z$ are Jacobi's eta, theta and zeta functions with modulus $m$, respectively. This wave function gives a dispersion relation $\omega=\mbox{cn}(s_0,m)\mbox{dn}(s_0,m)/\sqrt{m}$. To ensure the crystal momentum $q=iZ(s_0,m)/\sqrt{m}$ is real, we have two types of choice for the constant $s_0$, $s_0=i\alpha$ or $s_0=K(m)+i\alpha$, with $\alpha$ a real number confined to the first Brillouin zone, $-K(m_1)<\alpha<K(m_1)$. It is interesting to note that, generally, the solution given by Eq. (\ref{psi}) acquires a phase change of $2K(m)Z(s_0)$ after travelling through the DW, while keeping its amplitude unchanged. This conclusion is similar to that obtained for a DW with Walker profile \cite{Yan11}.

The frequency given by the choice $s_0=K(m)+i\alpha$ is $\omega=im_1 \mbox{sn}(\alpha,m_1) \mbox{cn}(\alpha,m_1) /\mbox{dn}^2 (\alpha,m_1)$, which is a pure imaginary number. Actually, any complex $s_0$ with a real part will give rise to a complex frequency with an imaginary part. As pure imaginary frequency corresponds to zero energy excitation with finite life time, $s_0 = K(m) + i\alpha$ does not generate SW eigenmodes. In addition,  in the limit $l\rightarrow\infty$, both the crystal momentum and the oscillation amplitude approach zero, indicating uniform and localized excitation profile. Given those considerations, we will not consider the choice $s_0=K(m)+i\alpha$ for the interaction between SWs and DWs in the following.

\begin{figure}
\onefigure[width=0.9\linewidth]{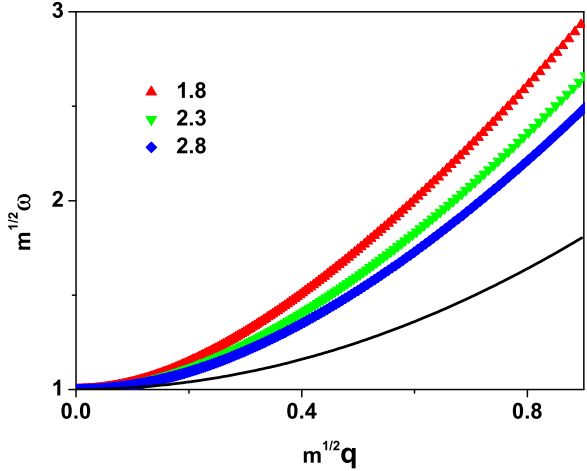}
\caption{Dispersion relation of magnons in a DW with half length $l$, whose value is shown in the legend. The solid line corresponds to the dispersion relation of magnons in a single domain state.}
\label{f2}
\end{figure}

For the choice $s_0=i\alpha$, the eigenfrequency is given by $\omega = \mbox{dn}(\alpha,m_1)/\sqrt{m}\mbox{cn}^2(\alpha,m_1)$. The corresponding crystal momentum reduces to $q\sqrt{m} = \alpha \pi /2K(m)K(m_1)- p\sqrt{m} +Z(\alpha,m_1)$ with $p=\mbox{sn}(\alpha,m_1) \mbox{dn}(\alpha,m_1)/ \sqrt{m} \mbox{cn} (\alpha,m_1)$. In Fig. \ref{f2}, $\omega$ versus $q$ curves corresponding to several $l$ are displayed. In the limit $m\rightarrow1$, the dispersion relation reduces to $\omega = 1+k^2$ ($k=\tan\alpha$ is the wave number), whose eigenfunction is $\psi= (\tanh x-i k)\exp(i k x)/(1-i k)$. As already noticed in literature \cite{Yan11,Winter61,Thiele73}, for this eigenfunction, the incoming and outgoing waves have the same amplitude, hence there is no reflection of magnons in the DW. This conclusion remains true for any finite value of $l$. However, for magnons incident on the finite DW from outside, due to the mismatch of dispersion relations of magnons inside and outside the DW, part of the incident wave will be reflected. That is to say, the DW potential is not reflectionless anymore for the incoming SW with a different dispersion relation.

\begin{figure}
\onefigure[width=0.9\linewidth]{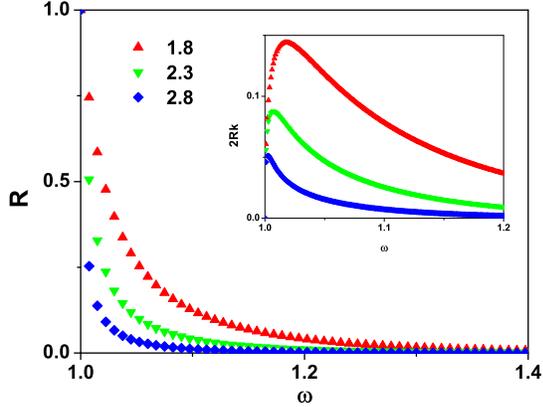}
\caption{Reflection coefficient $R$ as a function of frequency $\omega$, for a SW incident upon a DW with half length $l$, whose value is shown in the legend. Inset shows the normalized momentum transfer to the DW, $2Rk$, due to the reflection of the incident SW.}
\label{f3}
\end{figure}

To get the reflection property of the DW, suppose there is a SW $\psi=\rho\exp(ikx)$ incident from $-l$, with amplitude $\rho$ and dispersion relation $\omega = 1+k^2$. Evanescent SWs with imaginary $k$ are not considered, since they carry no momentum. For convenience, the whole 1D $x$ space is divided into three regions, with region I left to the DW, region III right to the DW, and the DW locating in region II. Then the SWs in these three regions can be written as
$\psi_I=\rho e^{ikx} + r \rho e^{-ikx}$, $\psi_{II}=a \rho \psi_+ + b \rho \psi_-$ and $\psi_{III} = t \rho e^{ikx}$, after reflection and transmission of the incident SW. $r$ and $t$ are amplitude reflection and transmission coefficients, and $\psi_\pm$ are right- and left-travelling eigenfunctions in the DW,
\begin{equation}
\psi_\pm=\frac{H(s\pm s_0)}{\Theta(s)}e^{\mp sZ(s_0)}.
\end{equation}
By requiring that the wave amplitude and its first derivative are continuous across the two boundaries of the DW, $\psi_I |_{x=-l} = \psi_{II} |_{x=-l}$, $\partial_x\psi_I|_{x=-l}=\partial_x\psi_{II}|_{x=-l}$, $\psi_{II}|_{x=l}=\psi_{III}|_{x=l}$, and $\partial_x \psi_{II} |_{x=l} = \partial_x \psi_{III}|_{x=l}$, the reflection coefficient can be obtained as
\begin{equation}
R=|r|^2=\frac{(k^2-p^2)^2\sin^2(2 q K(m))}{4k^2p^2+(k^2-p^2)^2\sin^2(2 q K(m))}.
\label{r}
\end{equation}
The boundary conditions used to get $R$ also ensure the continuity of probability. In Fig. \ref{f3}, $R$ for different values of $l$ is shown. It can be seen that $R$ is not zero for a finite range of the incidence frequency, which is in qualitative agreement with a previous analytical investigation based on the 1D Heisenberg model \cite{Liu79}. A similar conclusion was also reached in a micromagnetic study on the reflectivity of SWs by N\'{e}el walls \cite{Macke10}.

Formally, the reflectivity $R$ given by Eq. (\ref{r}) is identical to that for a non-relativistic electron traversing a potential barrier with hight $\omega_c = 1/\sqrt{m}$ and width $2K(m)$, with $k$ and $p$ acting as the momenta outside and inside the barrier, respectively \cite{barrier}. This coincidence is not accidental, since the same Schr\"{o}dinger equation is used to describe the particles, either magnons or electrons, in both cases. When the incident $\omega$ is less than $\omega_c$, SWs in the DW become evanescent and the incident SW mostly gets reflected. Above $\omega_c$, SWs can propagate in the DW, hence the incident SW mostly transmits across the DW, resulting in a decrease of $R$ upon increase of $\omega$. The transition at $\omega_c$ gives a corresponding critical wavelength $\lambda_c = 2\pi\delta\sqrt{l}/\sqrt{K(m)-l}$, in terms of the DW width $\delta$, for the incident SW. For SWs with wavelength $\lambda$ shorter than $\lambda_c$, the reflectivity $R$ is considerably smaller than unity. This quantify the intuitive argument used in Ref. \cite{XWang12}: If $\lambda$ is small, the variation of the magnetization in the DW can be neglected on the length scale of $\lambda$, inducing no significant perturbation to the SW propagation and negligible reflectivity. However, in the large $\lambda$ limit, the SW resolves the rotation of magnetization caused by the presence of the DW during its propagation and gets reflected. As is obvious from the definition of $p$ and $\omega$, in the limit $l\rightarrow\infty$, $p\rightarrow k$, $\omega\rightarrow 1 + k^2$ and the reflectivity is asymptotically zero, which confirms that the mismatch of dispersion relations is a prerequisite for reflection of SWs.

The finite reflectivity of SW implies that, upon reflection and transmission, the momentum of the incident SW is changed from $\rho^2\hbar k$ to $(-R+T)\rho^2\hbar k$, where $\hbar$ is the Planck constant divided by $2\pi$ and $T=1-R$ is the transmission coefficient. The net change in momentum thus is given by$(-R+T-1)\rho^2\hbar k=-2R\rho^2\hbar k$. The corresponding MT to the DW is given by $2R\rho^2\hbar k$, provided the total momentum is conserved. Since the MT is a product of two factors, $R$ and $k$, which are decreasing and increasing functions of frequency respectively, the competition between them brings about peaks shown in the inset of Fig. \ref{f3}. The position of those peaks depends on $l$, and the peak position shifts to higher frequency values with decreasing $l$. For small $l$, the peak is very broad, signifying that the MT can remain finite for a large range of frequency. Hence, for the DW confined in 1D space, the consequent counteraction force due to the MT can play an important role in DWM.

The angular momentum transfer from SWs to the DW is determined by the spin current \cite{Yan11} $j_s \propto (1-R) (\psi\partial_x\psi^* - \psi^* \partial_x \psi)$. The speed of the DW driven by this magnonic STT is $v_s=-(1-R)\rho^2v_g/2$, where $v_g = \partial\omega/\partial k$ is the group velocity of the incident SW. In contrast, the linear MT is determined by the momentum current $j_p \propto R\hbar k (\psi\partial_x\psi^* - \psi^*\partial_x\psi)$. The speed of the DW due to the MT is $v_p=2\rho^2 R \hbar k/m_D a^2$, where $m_D=2 E(m)/\sqrt{m}\mu_0\gamma^2\delta$ is the D\"{o}ring mass \cite{Doring48,Janak64} and $E(m)$ is the Jacobi elliptic integral of the second kind, whose value varies between $E(0)=\pi/2$ to $E(1)=1$. $v_p$ thus obtained is of the order of the speed driven by the magnonic STT, which can be seen by considering the ratio $v_p/|v_s| = \sqrt{m}R/E(m)(1-R) \times \hbar \gamma \mu_0M_s/2a^2\sqrt{KA}$. For $a$ = 2 {\AA}, $M_s$ = 1 $\mu_B/a^3$, $K$ = 10$^5$ J/m$^3$ and $A$ = 10 pJ/m, this ratio is $\sim$ 0.4$\sqrt{m}R/E(m)(1-R)$. Hence, for DWM driven by SWs, in the vicinity of the spin-wave cutoff frequency, the MT effect can be comparable in magnitude to the magnonic STT, since there the wavelength of the incoming SW is large and the reflection coefficient is finite (Fig. \ref{f3}). Further, very close to the spin-wave cutoff frequency ($\omega\sim 1$ and $R\sim 1$), the MT effect is more conspicuous than the magnonic STT effect. Needless to say, this is only a rough comparison of order for the two effects in the simplest 1D case considered here, neglecting magnetostatic interaction. In higher dimensions, more modes will emerge and modify the actual terminating speed obtainable, even in the absence of magnetic damping.

Experimental realization of DWM by magnonic MT can be achieved through various methods. The most direct one is to apply a GHz magnetic field perpendicular to the anisotropy direction, in a configuration identical to that used in conventional ferromagnetic resonance experiments. The only difference here is the requirement to apply the magnetic field locally, facilitating local injection of SWs with definite frequency and momentum. To do this, a coplanar waveguide can be put on top of a ferromagnetic nanowire \cite{Vlaminck08} with perpendicular anisotropy, in which a Bloch DW is nucleated. Besides GHz magnetic field, local injection of high density current \cite{Sekiguchi12} can also initiate magnetization oscillation through the STT effect. Elastic waves \cite{Weiler11} and spin-orbit field \cite{DWang12} in properly engineered material systems can serve as alternative means to locally exciting SWs that subsequently propagate to activate DWM.

To conclude, the reflection of magnons incident upon a domain wall confined in space is studied by considering the excitation spectrum of the domain wall, using 1D continuum micromagnetic analysis. Finite reflection coefficient is found in the vicinity of the spin-wave cutoff frequency. The momentum transfer from the incident magnons due to this finite probability of reflection can exert a counteraction force on the domain wall, which in turn can be utilized to drive domain wall motion. In stark contrast to the magnonic spin transfer torque, which moves a domain wall opposite to the propagation direction of the spin waves, the linear momentum transfer induced domain wall motion is in the direction of the spin wave motion.

\acknowledgments
X. W. and G. G. acknowledge financial support from the National Natural Science Foundation of China under Grant No. 60571043, Doctoral Fund of Ministry of Education of China under Grant No. 20120162110020, and the Scientific Plan Project of Hunan Province of China under Grant No. 2011FJ3193.

\end{document}